\documentstyle[amssymb,preprint,aps]{revtex}
\tightenlines

\begin{document}
\title{Spatio-Temporal Scaling of Solar Surface Flows}
\author{J. K. Lawrence and A. C. Cadavid}
\address{Department of Physics and Astronomy\\
California State University, Northridge}
\author{A. Ruzmaikin}
\address{Jet Propulsion Laboratory\\
California Institute of Technology}
\author{T. E. Berger}
\address{Lockheed Martin Space and Astrophysics Laboratory}
\date{\today }
\maketitle

\begin{abstract}
The Sun provides an excellent natural laboratory for nonlinear phenomena. We
use motions of magnetic bright points on the solar surface, at the smallest
scales yet observed, to study the small scale dynamics of the photospheric
plasma. The paths of the bright points are analyzed within a continuous time
random walk framework. Their spatial and temporal scaling suggest that the
observed motions are the walks of imperfectly correlated tracers on a
turbulent fluid flow in the lanes between granular convection cells.
\end{abstract}


\preprint{HEP/123-qed}

\narrowtext

In the outer 30\% of its radius the Sun's plasma is convectively unstable,
and thermonuclear energy from the core is transported by overturning motion
to the visible photosphere. The convective motions are most readily seen as
the photospheric pattern of bright ``granules'' of hot, rising gas,
surrounded by a network of ``intergranular lanes'' of cooler, darker,
downflowing gas. The horizontal extent of a granular cell is $\sim $1000 km
with a range of 200 km to 2500 km. The typical separation of granule
centroids is 700 km; typical lifetimes are 5 - 15 min.

Magnetic flux threading the solar surface is not uniformly distributed. By
processes which are not well understood, it is concentrated into discrete
kilogauss fibrils of diameter $\lesssim $ 150 km. These accumulate in the
intergranular lanes and vertices where they disturb the radiative properties
of the photosphere and become apparent to observers through spectral line
emission. Such magnetic bright points (MBPs) serve as tracers of the
evolution of the granular flow field and of small scale motions in the
intergranular network.

With Reynolds number $\approx 10^{7}$ turbulent flows are expected in the
intergranular lanes. Numerical simulations of convection near the solar
surface \cite{steinnordlund98} indicate that turbulent vortices are formed
in the intergranular lanes where the outflows from neighboring granules
collide and turn downward. Observationally, spectrograms taken in
magnetically and temperature insensitive spectral lines show line broadening
in excess of known Doppler contributions, indicating the presence of
turbulent flows on the borders of granules \cite{nesisetal93}. These
observations cannot rule out other types of motion. However, the unique,
high-resolution data set described below does provide access to the actual
dynamics of the flows and reveals some of their quantitative characteristics.

Here we study subgranular motions at the highest resolution yet achieved 
\cite{bergeretal98b,lofdahletal98,bergeretal98a}. The data were taken on
October 5, 1995 with the 50 cm Swedish Vacuum Solar Telescope on La Palma,
Spain and consist of two cospatial and simultaneous (within 10 ms){\em \ }%
time series of 178 digital image pairs spanning 71 minutes. One set of
images was taken through a filter passing wavelengths $430.5\pm 0.5$ nm
located in the G-band emission of the CH radical. The other set of images
was made through a wide-band filter passing wavelengths $468.6\pm 5.0$ nm
emphasizing the background continuum emission. A 21 Mm $\times $ 21 Mm area
of enhanced magnetic activity near the center of the solar disk was imaged,
containing hundreds of MBPs. Exposure time was 20 ms, and the pixel scale in
the digital images was 60 km on the Sun. The images were checked in real
time for seeing sharpness, and the best image pair in each 25 s interval was
recorded. Both time series were restored to near the telescope diffraction
limit (0.2 arcsec) by phase diversity reconstruction \cite{lofdahletal98}.
The final resolution is 165 km in the G-band and 180 km in the broad-band
channel. The images were further processed to remove instrumental effects
and geometric distortion due to atmospheric seeing.

Next, the wideband images were subtracted from their G-band counterparts,
removing most of the granulation pattern common to both sets. Finally the
images were subjected to a high-pass spatial filter to emphasize small-scale
features \cite{bergeretal95}, and a threshold was imposed to identify MBPs.
The result is a three dimensional binary (MBP or not) array with two 350
pixel axes corresponding to x and y spatial directions and a third axis of
178 pixels corresponding to the time. A given MBP is tracked by segmentation
of its binary object tree from the data cube. The segmentation is
implemented using one-dimensional voxel addressing and searching over
limited volumes around a given voxel for valid neighbors. After segmentation
the individual tree is split into branches and the motion of each sub-object
is measured with respect to a common reference point.

This procedure delivers strings of x and y positions and times, at roughly
24 s intervals, for 1800 individual MBPs. The longest sequence for an
individual MBP object spans 71 min, the shortest 9 seconds. During
approximately 1 hr the MBPs typically cover an rms distance on the order of
a granular diameter. Such an MBP path is illustrated in Fig. 1. The MBPs
undergo significant morphological changes on a time scale of 100 s and are
continuously merging and splitting as a result of their interaction with the
granular flow \cite{bergertitle96}. Consequently, the data strings contain a
number of large jumps, many representing artifacts related to the merging
and splitting of two or more MBPs. For example, the trajectory shown in Fig.
1 contains four jumps with $v>c$; the largest is $v=14.6$ km s$^{-1}$. To
ensure their removal, we follow past practice \cite{bergeretal98b} and
employ a cutoff above the photospheric sound speed $c=7$ km s$^{-1}$.

Because of the discrete structuring of the magnetic flux, the motions of the
MBP tracers are naturally described in terms of random walks \cite
{leighton64}. We will use a general equation for the time evolution of the $%
q-th$ moment of the tracer displacements: 
\begin{equation}
<r^{q}>^{1/q}=D\cdot t^{\gamma (q)/2}
\end{equation}
where $r$ is the total displacement during time $t$, and an appropriately
weighted average is taken over all observed displacements. Fig. 2 shows $%
<r^{q}>^{1/q}$ as a function of $t$\ for two values of $q$. We see excellent
power law scaling over two orders of magnitude, up to 35 minutes in time.
This permits precise determinations of $\gamma (q)$, which are shown in Fig.
3. While Brownian walks give $\gamma (q)=1$ for all $q>-1$, we find that $%
\gamma (q)>1$ and varies with $q$. Although diffusion studies are usually
limited to mean square displacements, i. e. $q=2$, here the general $q$
dependence proves to be crucial for correct analysis of the motions.

Figs. 2 and 3 indicate that for the MBP motions $\gamma (2)=1.13\pm 0.01$
implying superdiffusion, and that in the limit of small $q$ we approach the
even more superdiffusive value $\gamma (0)$ $=1.27\pm 0.01$. Our analysis
below will indicate that in the ideal situation of infinite, precise data we
should find that $\gamma (q)$ is independent of $q$. The problem here is to
find that ideal value from limited, imperfect data. When we use no velocity
cutoff, a case not illustrated, we find $\gamma (2)<1$ indicating
subdiffusion, but again that $\gamma (0)\approx 1.3$. We will see below that
it is $\gamma (0)$, rather than $\gamma (2)$, that reliably characterizes
the MBP walks.

The superdiffusion we have found fits naturally into the framework of the
continuous time random walk (CTRW) approach \cite{klafteretal87}. This
method focuses on the statistical properties of the processes under
consideration. It handles superdiffusion by means of L\'{e}vy flights, with
an infinite mean squared displacement, by adding a time cost to the long
jumps, so that the mean squared displacement as a function of time
converges. In a broader context, previous experimental work \cite
{solomonetal93} has shown that tracer motion in a rotating flow presents
anomalous diffusion which can be analyzed in terms of L\'{e}vy flight
motions as we do here. More recent experimental work \cite{jullienetal99}
finds that the separations of particle pairs in a turbulent flow, in the
Richardson regime, show a stretched exponential form rather than the
L\'{e}vy power law form. This result, however, can be recovered from
particle pairs undergoing relative motions with an asymmetric L\'{e}vy
distribution \cite{sokolovetal00}.

Our use of the CTRW technique is based on two adjustable parameters. The
first, $\lambda $, is the exponent of a power law distribution $\sim
r^{-\lambda }$ of spatial steps of tracers due to underlying fluid motions.
The second, $\nu $, connects the spatial step lengths $r$ to elapsed times
according to $r\propto t^{\nu }$. By looking at the spatial scaling
properties of the tracer motions, we can directly estimate these parameters
for our case. Fig. 4 shows the distribution of lengths for all solar MBP
steps between consecutive images excluding those with $v>c$. A fit for all $%
r\geq $ $40$ km gives $\lambda =2.65\pm 0.13\approx 8/3$. The assumed step
time scaling gives a velocity $v=r/t\propto r^{1-1/\nu }$. We estimate this
exponent by calculating the velocities of all solar MBP steps with $t\leq
1000$ s, but rejecting those with $v>c$. A plot of $v$ versus $r$ is shown
in Fig. 5. The best fit for $20$ km $\leq r\leq 1000$ km gives $1-1/\nu
=0.35\pm 0.03\approx 1/3$. Thus we find $v\propto r^{1/3}$, leading to $\nu
\approx 1.5\pm 0.2$ and implying Kolmogorov ``K41'' scaling with $%
r^{2}\propto t^{3}$ \cite{kolmogorov41,frisch95}. However, we have found
above that the MBP random walks give $<r^{2}>\propto t^{\gamma }$ with $%
\gamma $ superdiffusive but less than $3$. A key result of the CTRW approach
is that, because the tracers may be imperfectly correlated with the
underlying fluid flow, ``Kolmogorov scaling does not necessarily imply
Richardson's law \cite{shlesingeretal87}.''

To connect our notation with that common in the CTRW literature \cite
{klafteretal87,shlesingeretal87,zumofenetal89} we note that the formalism is
based on a probability density function $\psi (r,t)=Cr^{-\mu }\delta
(r-t^{\nu })$. Our parameter $\lambda =\mu -1/\nu -(d-2)$. In the present
case the dimension $d=2$. The $1/\nu $ term arises from the measure
associated with the delta function argument.

The CTRW technique allows analytical calculation of $<r^{q}>^{1/q}\propto
t^{\gamma (q)/2}$ for $q=2$. The time evolution of the moments, that is, the
value of $\gamma $, is strongly dependent on the values of $\lambda $ and $%
\nu $. There are four different cases depending on the degree to which the
tracers we follow are coupled to the fluid motions. These are illustrated in
Fig. 6 as regions in the two dimensional $\lambda $, $\nu $ plane. In Case I
with $\lambda \geq 3$ and $\nu \geq 1/(\lambda -1)$ one finds $\gamma (2)=1$%
. This corresponds to a brief correlation time $\tau $ of the tracers with
the flow and thus normal Brownian diffusion. In Case II $\lambda >3$ but $%
\nu <1/(\lambda -1)$. This corresponds to strong coupling with $\tau =\infty 
$ and $\gamma (2)=\nu (\lambda -1)<1$ giving subdiffusion. In Case III $%
\lambda <3$ and $\nu >1/(\lambda -1)$. In this intermediate case $\tau $ is
finite but includes significant coupling of the tracers with the flow. We
find $\gamma (2)=1-\nu (\lambda -3)>1$ and superdiffusion. Finally, in Case
IV we have $1\leq \lambda <3$ and $\nu \leq 1/(\lambda -1)$. Now $\tau
=\infty $, and the tracers are completely coupled to the flow. We find $%
\gamma (2)=2\nu $. This is superdiffusive when $\nu >1/2$.

The analytical calculation of $\gamma (2)$ is verified by numerical
simulation \cite{zumofenetal89}. We extend the simulations to arbitrary
order in the range $0.1\leq q\leq 4.0$ by adopting a simple, repetitive
algorithm for generalized random walks: step in a random direction by a
distance $r$ chosen with a probability $\varpropto r^{-\lambda }$ while
advancing the time by $t\varpropto r^{1/\nu }$.

In the limit of very many trials we find that $\gamma (q)\rightarrow \gamma
=const$, which differs from the decreasing $\gamma (q)$ shown by the solar
MBPs in Fig.3. However, the data set is limited in time and number of
tracers so the statistics are important. Let $\widetilde{\gamma }(q)$
represent the diffusion exponent calculated by averaging over 100 walks
simulated with our CTRW algorithm, roughly matching the observational
situation. An average over 1000 such $\widetilde{\gamma }(q)$ gives a
constant $<\widetilde{\gamma }(q)>\approx 1.3$, as illustrated in Fig. 3 for
a simulation with $\lambda =8/3$ chosen to match the solar data. Also shown
in Fig. 3 are the one standard deviation limits of the 1000 runs. These
indicate the range within which one expects to find a single realization of $%
\widetilde{\gamma }(q)$. Unless $q\ll 1$, the form of an individual $%
\widetilde{\gamma }(q)$ is dominated by a small number of\ the largest
jumps. Depending on their timing in individual runs they can give large
fluctuations in, for example, $\widetilde{\gamma }(2)$. For $q\ll 1$ the
values of $\widetilde{\gamma }(q)$ are determined more equally by all jumps
and characterize the asymptotic mean of $\gamma (q)$ for all $q$. The
distribution of the underlying $\widetilde{\gamma }(q)$ is narrow for $%
q\lesssim 0.5$ permitting a robust measurement.

The values of $\lambda $ and $\nu $ determined from the scaling of the solar
MBP space and time steps are shown on the plot of Fig. 6. These place the
solar flows on the $\nu =3/2$ line corresponding to Kolmogorov scaling.
However, the value of $\lambda $ selects the superdiffusive intermediate
regime between Brownian and Richardson diffusion. Thus, the MBPs are
imperfect tracers of the flow. For the Richardson case with $\gamma =3$ we
would need perfect correlation of the MBPs with the flow. Instead the
parameters predict a diffusion exponent $\gamma =1.54\pm 0.39$. From the
simulations based on the same $\lambda $ and $\nu $ we find $\gamma
(0.1)=1.28\pm 0.05$, and $\gamma (2)=1.24\pm 0.33$; the uncertainties show
why using $q\rightarrow 0$ is preferable to using $q=2$ when dealing with
data. By direct measurement of the random walks we found $\gamma (0)=1.27\pm
0.01$and $\gamma (2)=1.13\pm 0.01$. Mindful of the limited data set, we
adopt the larger value. The result $\gamma \approx 1.3$ holds independent of
any velocity cutoff.

It is no surprise that the MBPs are imperfect tracers of the fluid flows.
They are not truly local objects, but rather the intersections with the
photosphere of extended flux tubes which are influenced by various motions
at any given time and are restricted to intergranular lanes. Nevertheless,
the CTRW approach to anomalous diffusion via L\'{e}vy flights reveals
quantitative properties of the small scale photospheric flows and their
coupling to the magnetic flux tubes. The spatio-temporal scaling of the MBP
dynamics indeed indicates the presence of turbulence in intergranular lanes.
Our quantifying of the deviation of the MBP motions from the underlying
fluid motion should contribute to a better physical understanding of flux
tubes and the solar medium.

This work was supported in part by NSF Grants ATM-9628882 and ATM-9987305.
T. Berger was supported by NASA SR\&T contract NASW-98008. The SVST is
operated by the Swedish Royal Academy of Sciences at the Spanish
Observatorio del Roque de los Muchachos of the Instituto de Astrofisica de
Canarias.

\begin{figure}[tbp]
\caption{Typical solar MBP trajectory. Distances are in Megameters.
Positions were observed at intervals of about 24 s; the total elapsed time
is 50 min.}
\label{Fig1.eps}
\end{figure}

\begin{figure}[tbp]
\caption{Moments $<r^{q}>^{1/q}$ of the cumulative solar MBP displacements $%
r $ in Megameters versus time in minutes for a sonic velocity cutoff and
moment orders $q=0.1$ (filled circles) and $q=2$ (open circles). If error
bars are not visible, they are smaller than the symbols.}
\label{Fig2.eps}
\end{figure}

\begin{figure}[tbp]
\caption{The multiscaling exponent $\protect\gamma (q)$ versus moment order
q. Solid line: solar MBPs with a sonic cutoff $v\leq c$, open circles:
continuous time random walk (CTRW)mean value for 1000 runs of 100 walks
each, dotted lines: 1 standard deviation limits of the CTRW case indicating
the expected range for a single 100 walk realization.}
\label{Fig3.eps}
\end{figure}

\begin{figure}[tbp]
\caption{Filled circles: Logarithms of numbers of step lengths in 5 km bins
versus the logarithm of step lengths. The solid line has slope -8/3 and
represents a fit for all $r>40$ km.}
\label{Fig4.eps}
\end{figure}

\begin{figure}[tbp]
\caption{Filled circles: Step velocities of MBP motions versus step length
for all steps occurring over 1000 s or less and with $v\leq 7$ km s$^{-1}$.
Error bars are smaller than the symbols. Dashed line: best linear fit for
step lengths between $20$ km and $1000$ km. The slope is $0.35\pm
0.03\approx 1/3$.}
\label{Fig5.eps}
\end{figure}

\begin{figure}[tbp]
\caption{The two dimensional $\protect\lambda $, $\protect\nu $ parameter
space. The solid lines $\protect\lambda =3$ and $\protect\nu =1/(\protect%
\lambda -3)$ divide the space into four regions with differing correlation
properties of the tracers with the underlying flows and differing diffusion
exponents. The dotted line at $\protect\nu =3/2$ indicates Kolmogorov
scaling. The heavier dotted line at $\protect\nu =3/2$ and $\protect\lambda %
<5/3$ indicates the locus of Richardson diffusion with $\protect\gamma =3$.
The solar symbol represents the values of $\protect\lambda $ and $\protect%
\mu $ estimated directly from the solar MBP walks.}
\label{Fig6.eps}
\end{figure}

\bigskip

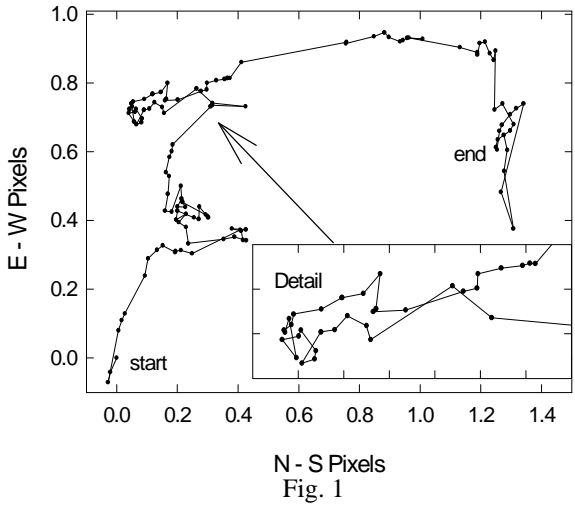
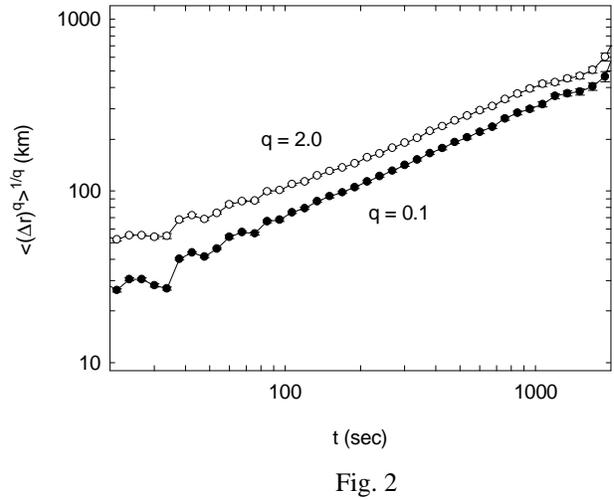
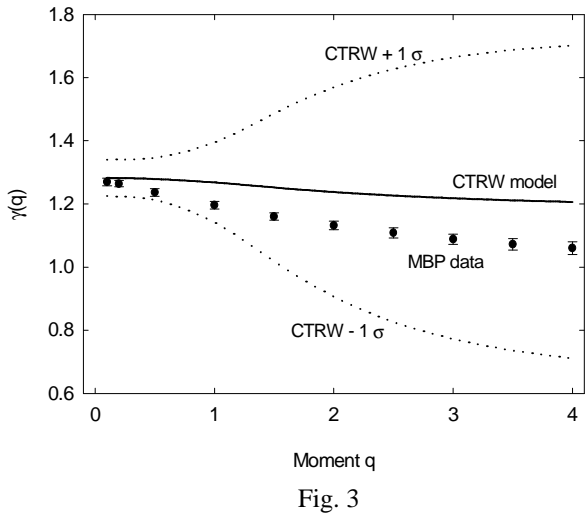
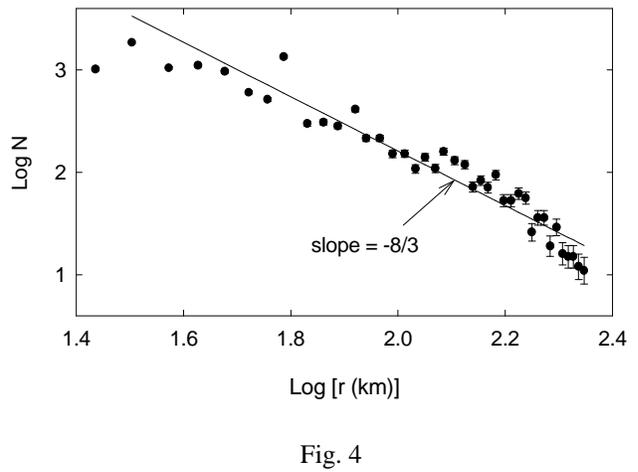
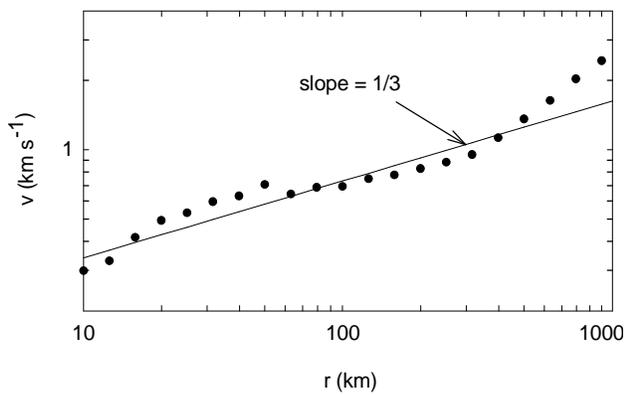
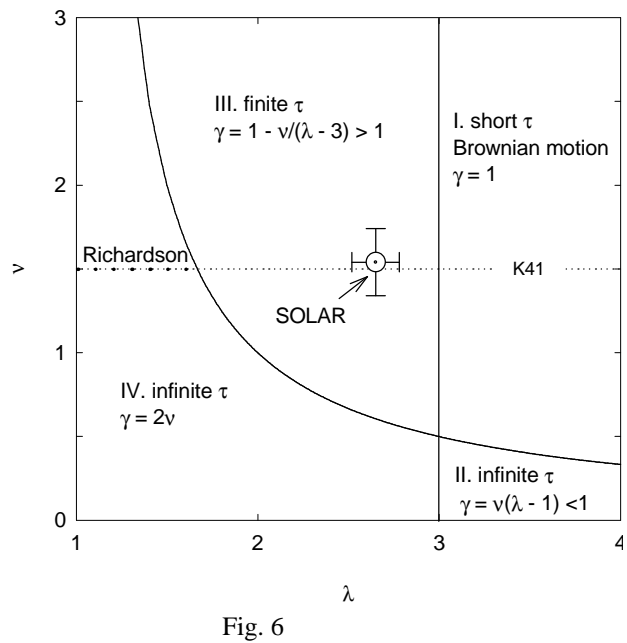
\end{document}